\documentstyle[draft]{mn}

\newcommand{\ltappeq}{\raisebox{-0.6ex}{$\,\stackrel
{\raisebox{-.2ex}{$\textstyle <$}}{\sim}\,$}}
\newcommand{\gtappeq}{\raisebox{-0.6ex}{$\,\stackrel
{\raisebox{-.2ex}{$\textstyle >$}}{\sim}\,$}}

\title[Jet Features in Cataclysmic Variables]{On jet features in the  optical spectra of cataclysmic variables}
\author[C. Knigge \& M. Livio]{Christian Knigge and Mario Livio\\Space Telescope Science Institute, 3700 San Martin Drive, Baltimore, MD 21218\\knigge@stsci.edu, mlivio@stsci.edu}

\date{Accepted ????.
      Received ????;
      in original form ????}
 
\pagerange{\pageref{firstpage}--\pageref{lastpage}}
\pubyear{1997}
 
\begin{document}
 
\maketitle
 
\label{firstpage}
 
\begin{abstract}

Blue- and red-shifted Hydrogen and Helium satellite 
recombination lines have recently been discovered in the optical 
spectra of at least two supersoft X-ray sources (SSSs), RX~J0513-069 
and RX~J0019.8+2156, and tentatively also in one short-period 
cataclysmic variable star (CV), the recurrent nova 
T~Pyx. These features are thought to provide 
evidence for the presence of highly collimated jets in these
systems. No similar spectral signatures have been detected in the
spectra of other 
short-period cataclysmic variables, despite a wealth of existing 
optical data on these systems. Here, we ask if this 
apparent absence of ``jet lines'' in the spectra of most CVs
already implies the absence of jets of the kind that appear to be
present in the SSSs and perhaps T~Pyx, or whether the current lack of 
jet detections in CVs can still be ascribed to observational
difficulties.

To answer this question, we derive a simple, approximate scaling
relation between the expected equivalent widths of the observed jet
lines in both types of systems and  the accretion rate through the
disk, $EW(line) \propto  \dot{M}_{acc}^{\frac{4}{3}}$. We 
use this relation to predict the strength of jet lines in the spectra
of ``ordinary'' CVs, i.e. systems characterized by somewhat lower
accretion rates than T~Pyx. Making the assumption that the features
seen in T~Pyx are indeed jet lines and using this system as a
reference point, we find that if jets are present in many
CVs, they may be expected to produce optical satellite recombination
lines with EWs of a few hundredths to a few tenths of Angstroms in
suitably selected systems. A similar prediction is obtained if the 
SSS RX~J0513-069 is used as a reference point. 
Such equivalent widths are small enough to account for the
non-detection of jet features in CVs to date, but large enough to
allow them to be detected in data of sufficiently high quality, if
they exist.

\end{abstract}


\section{Introduction}
\label{introduction}

One of the most intruiging empirical connections that has emerged from
observations of accretion-powered objects on all astrophysical scales
is that between accretion disks and powerful bipolar outflows and
jets. Indeed, the occurrence of mass loss in this form has been 
established in members of essentially all classes of 
(presumed) disk-accretors, including close binary systems such as
X-ray binaries, supersoft X-ray 
sources (SSSs) and non-magnetic cataclysmic variable stars (CVs),
young stellar objects such as T-Tauri and FU~Orionis stars, and 
active galactic nuclei and quasars.  However, a complete theoretical
understanding of this empirical disk-wind (or disk-jet) connection has
not yet been achieved. Recent reviews of the subject with references
to the relevant observational literature may be found in Livio
\shortcite{livio4,livio5}.

A potentially vital clue to the origin of mass loss from accretion
disk systems is provided by the fact that in at least some members of 
all but one class of objects the outflow collimation appears to be 
very tight, with half-opening angles, $\theta_{max}$, of 
no more than a few degrees (we will refer to such flows as {\em jets} 
hereafter). The sole exception to this rule are CVs (see however below). 
In these systems, the presence of mass loss has nevertheless been
clearly established, based on the shapes and eclipse behavior of the
ultraviolet (UV) resonance lines (e.g. Drew 1991),
but the inferred collimation of the corresponding outflows is weak
($\theta_{max} \gtappeq 45^o$; Shlosman, Vitello \& Mauche 1997;
Knigge \& Drew 1997).

The apparent absence of jets in CVs may hold the key to an improved
theoretical understanding of the disk-wind connection. If confirmed, any
``generic'' model (in the sense of being applicable to more than a
particular class of objects) for the origin of mass loss from accretion
disks must be able to explain why this mass loss takes the form of
jets in all systems but CVs. This would be quite a restrictive
constraint, particularly in the light of two recent observational
developments.

The first of these is the detection of blue- and red-shifted 
satellite emission features to the optical Hydrogen and Helium 
recombination lines in the SSSs RX~J0513-069, RX~J0019.8+2156 
and (possibly) CAL~83
\cite{crampton1,crampton2,southwell1,southwell2,becker1}. These
features cannot be attributed to any other ionic species and are
therefore thought to arise in some kind of bipolar
outflow. The combination of small width and large displacement 
from line center exhibited by these satellite lines further suggests
that the corresponding outflows are very highly collimated, i.e. 
that they are jets (c.f. the prototypical jet system SS~433;
Vermeulen~1993).

If this identification is correct (which we will assume throughout 
this paper) it is significant, because SSSs and CVs are extremely
similar types of objects: both are semi-detached binary 
systems in which a Roche-lobe filling secondary star transfers material
via an accretion disk onto a white dwarf (WD) primary. The main 
difference between SSSs and CVs is thought to be the rate at which this
mass transfer proceeds. In SSSs, the 
accretion rate is believed to be high enough ($\dot{M}_{acc} \sim 10^{-7}
- 10^{-6}$~M$_{\sun}$~yr$^{-1}$) to fuel steady nuclear hydrogen burning
on the surface of the WD (e.g. van den Heuvel et a. 1992). By contrast, 
the highest accretion rates 
encountered in CVs are about one order of magnitude lower than this
($\dot{M}_{acc} \sim 10^{-8}$~M$_{\sun}$~yr$^{-1}$) and 
thus insufficient to initiate steady nuclear burning on the
WD. Instead, the accretion can result in shell flashes which are
responsible for nova outbursts. This 
difference is one of the factors that led Livio (1997a) to
propose that the formation of powerful jets (as opposed to more 
weakly collimated bipolar outflows) may {\em require} the presence 
of an additional wind/energy source at the center of the accretion
disk.

The second recent observational finding of significance in
the present context is the tentative detection of similar
H$\alpha$~satellite lines 
in the optical spectrum of {\em one} CV, the recurrent nova T~{Pyx}
\cite{shahbaz1}. It is important to stress that no similar features 
have ever been detected in any other short-period CV, despite the 
fact that optical spectra of high enough 
quality to detect satellite lines of similar strength as in T~Pyx
(which, on average, have an equivalent width of about 1~\AA) should
be available for many of them. While it is possible that in some cases
they might have been overlooked (previous studies would not have
expected to see such features), it would nevertheless appear that
satellite lines of the strength seen in T~Pyx are not a common feature 
among CVs. For example, Shahbaz et al. (1997) did not detect similar
satellite lines in the spectrum of another recurrent nova,
U~Sco. (It should be noted that one-sided satellite lines have been
seen in the spectra of a few CVs, e.g. S193, V795~Her, BT~Mon [Szkody
1995; Haswell et al. 1994; Seitter 1984]; however, these probably
arise in other types of high velocity flows in these systems.)

The physical similarities between CVs and SSSs, on the one hand, and
the observational similarity between the satellite features in T~Pyx 
and those in SSSs such as RX~J0513-069, on the other, suggest
that T~Pyx may also harbor a well collimated jet. A fundamental
assumption we adopt in the present paper is that this is indeed the 
case. Note that, under this assumption, we would expect T~Pyx's binary
inclination to be higher than that of the SSSs listed above, since in
T~Pyx the ratio of satellite line widths to their displacements from
line center is somewhat smaller. A high inclination for T~Pyx ($i \sim
70^o$) is also indicated if one demands that the displacement of the
satellite features from line center 
should roughly correspond to the escape velocity from the WD, as
expected if they are formed in a jet (c.f. Shahbaz et al. 1997; Livio
1997a). It is acknowledged, however, that Shahbaz et al. (1997) 
also derived an inclination estimate from the peak-to-peak separation 
of the double-peaked H$\alpha$ line core, which, by contrast, turns
out to be very low ($i \sim 10^o$). While this estimate should be
regarded as a lower limit \cite{shahbaz1}, the case for a well
collimated jet in T~Pyx would have to be critically reexamined if 
the inclination of this system were shown to be $\ltappeq 50^o$ in 
the future. The analytic scaling relation derived in Section~2 would
of course nevertheless be valid and, we believe, useful, even if 
T~Pyx should eventually turn out not to contain a collimated jet.

Actually, the existence of a jet in T~Pyx is not entirely 
unexpected, since it is thought that recurrent novae in general, 
and T~Pyx in particular, are characterized by accretion rates that 
are higher (i.e. $\dot{M}_{acc} \gtappeq
10^{-8}$~M$_{\sun}$~yr$^{-1}$) than 
those in other types of CVs, such as dwarf-novae (DNe) and nova-like 
variables (NLs). In fact, Webbink et al. (1987) have suggested that 
intermittent nuclear burning might take place on the surface of the 
WD in T~Pyx even during quiescence, i.e. between nova outbursts. 
If so, Livio's (1997a) hypothesis could be used to reconcile the
presence of jet lines in SSSs and T~Pyx with the absence of similar
features in the spectra of other CVs.

The main goal of the present paper is to determine whether such
reconciliation is actually required at present. Thus, we will ask
whether the apparent absence of jet lines in the existing optical
spectra of most CVs, particularly NLs and DNe in outburst
(i.e. systems in which an optically thick accretion disk is present
but no nuclear burning occurs) already implies the {\em absence} of 
jets of the kind seen in the SSSs and (possibly) T~Pyx, or whether the
current lack of jet detections in CVs can still be ascribed to
observational difficulties. In our attempt to answer this question, we
will take as our fundamental working hypothesis that the main 
difference between CVs and the SSSs (as well as perhaps T~Pyx) -- the
presence of an extremely hot WD at disk center -- is irrelevant to the 
formation of the observed jets. The corollary of this hypothesis is
that CVs must actually drive jets of the same kind that appear to be 
present in the SSS and T~Pyx. Our goal is to estimate the expected
equivalent widths of CV jet lines within the framework of this
hypothesis. In so doing, we will try to make as few references as
possible to specific models for the jet formation and line emission
mechanisms, and instead use only simple and general physical
arguments.

As already noted above, under our working hypothesis, jets akin to
those in the SSSs must actually be present in CVs. This is
not in direct conflict with the relatively wide outflow opening angles
that have been inferred for 
CV winds from modeling the UV resonance lines: these spectral features
probe only the near-disk regime of the outflow -- out to at most a few
hundred $R_{WD}$ -- leaving collimation at larger distances as a
distinct possibility. 

\section{A scaling relation for the strength of jet-formed satellite
recombination lines}
\label{scale}

The main difficulty in deciding whether jet lines should already have
been detected in CVs (if these systems do indeed drive jets), lies in
the fact that the emission mechanism responsible for producing the
observed features in T~Pyx and the SSSs has not yet been
established. Given the narrowness and distinct location of the 
lines in the spectra of these objects, the emitting region may well 
be a shock far downstream in the flow, perhaps at the point where
collimation occurs. However, an obvious selection effect is at work
here -- jet lines are much easier to detect if they are well separated
from the main emission line -- and thus even this identification is
far from secure. Our working hypothesis can therefore not be tested by
direct modeling of the proposed jets (which is in any case impossible
since the jet driving mechanism in SSSs and other systems is also not
precisely known). 

To overcome this difficulty, we take an extremely general approach to
the problem and rely as much as possible on the similarity between
T~Pyx and the SSS, on the one hand, and ``normal'' CVs on the
other. Thus we only derive a rough, simple scaling relation between 
the equivalent widths (EWs) of these features and the relevant
physical parameters of a given system, which 
we expect to be approximately valid regardless of the detailed flow
dynamics and line emission mechanism. Such generality comes at a 
price, of course. In particular, we will have to make some, 
hopefully plausible, simplifying assumptions to arrive at our scaling 
relation, and will only be able to use it to compare similar systems 
(i.e. systems sharing common jet driving and line emission mechanisms). 
Under our working hypothesis, the SSSs, T~Pyx and other CVs are 
similar in this sense.

We begin by exploring the dependence of jet line EWs on the mass
transfer rate through the disk, since this is likely to be the most
relevant parameter in this respect. Assuming the optical light of both
CVs and SSSs to be dominated by radiation from an optically thick,
steady-state accretion disk, we expect the optical continuum flux to
scale roughly as (e.g. Webbink et al. 1987)
\begin{equation}
F_{c,opt} \propto \dot{M}_{acc}^{\frac{2}{3}}.
\label{continuum1}
\end{equation}

This assumption is almost certainly valid for CVs in a state  of high
mass accretion rate, i.e. NLs and DNe in outburst: optical eclipse
mapping studies show that most high-state CVs do approximately follow
the expected disk temperature distribution $T_{eff}(R) \propto
R^{-3/4}$ that is the most fundamental prediction of steady-state
accretion disk theory (e.g. Horne \& 
Stiening 1985; Rutten et al. 1992, Baptista et al. 1995). And while
model disk spectra based on the same theory do not always match
observed spectra in detail (e.g. Knigge et al. 1997, 1998), the
corresponding discrepancies should be of minor imporance in the
present context.

To test whether Equation~\ref{continuum1} is also plausible for SSSs,
we have calculated the expected 
contribution of a very hot ($T_{eff}=500,000$~K), $1M_{\sun}$~WD to
the total optical light from such a
system, assuming an accretion rate through the disk of $\dot{M}_{acc}=
10^{-7} M_{\sun}$~yr$^{-1}$, a disk radius of $R_{disk}=50R_{WD}$ and
a representative inclination of $i=60^o$. Within the framework of 
the usual approximations -- that the disk is optically thick, and that 
it and the WD both emit as (ensembles of) blackbodies -- we find that
the WD contributes only about 2\% to the flux near $H\alpha$. For
larger disk radii and/or accretion rates, and for systems viewed 
at lower inclinations, the disk would dominate the visual flux even
more. Nevertheless, this number is almost certainly an underestimate, 
since our calculation neglected the possibility that some of the WD 
radiation might be reprocessed by the disk and the secondary. 
Because of this, we consider this assumption to be the weakest 
link in our chain of arguments when applied to SSSs and will return 
to it in Section~\ref{BLA3}.

Even if direct and reprocessed WD radiation are unimportant, 
Relation~(\ref{continuum1}) is only valid provided that 
\begin{equation}
T_{max} >> \frac{hc}{\lambda k} >> T_{out},
\label{limits}
\end{equation}
where $T_{max}$ is the maximum disk temperature, $T_{out}$ is the
temperature at the outer disk edge, $\lambda$ corresponds to the
optical waveband and all other symbols have their usual meaning. 
For CVs, this condition is usually satisfied approximately, though not
in detail. It is therefore useful to derive some upper
and lower limits on the power law index in Relation~(\ref{continuum1}) 
that are independent of this condition.

An upper limit can be derived by noting that the total disk luminosity
corresponds to one half of the accretion luminosity and thus scales as
$L_{disk} \propto \dot{M}_{acc}$. Given that the the peak of the disk
spectrum will lie shortward of the optical waveband for all reasonable
parameters, the steepest possible scaling of the optical continuum
flux with accretion rate, which corresponds to taking the spectral 
shape to remain constant as $\dot{M}_{acc}$ increases, is $F_{c,opt}
\propto \dot{M}_{acc}$. In reality, the spectrum will of course become
bluer as the disk becomes hotter with increasing $\dot{M}_{acc}$. 
A corresponding lower limit on the power law index in
Relation~(\ref{continuum1}) may therefore be derived by assuming that
the optical waveband lies on the Rayleigh-Jeans tail of the
spectrum. In this case, $F_{c,opt} \propto T \propto L_{disk}^{1/4}
\propto \dot{M}_{acc}^{1/4}$. Taking these limits into account, 
Relation~(\ref{continuum1}) becomes
\begin{equation}
F_{c,opt} \propto \dot{M}_{acc}^{{\left({\frac{2}{3}}\right)}^{+1/3}_{-5/12}}.
\label{continuum}
\end{equation}

It is worth stressing that our quoted ``errors'' on the power law
index in Relation~(\ref{continuum}) are quite conservative, in the 
sense that they are likely to bracket the dependence of the optical
flux on accretion rate for {\em any} plausible, optically thick model
of radiation liberated as a result of accretion. As a check, we
have constructed the function $F_{c,opt}(\dot{M}_{acc})$ directly 
from numerical models in which the accretion disk is assumed to be in
a steady-state and to radiate as an ensemble of blackbodies or stellar
atmospheres (see, for 
example, Knigge et al. 1997). We find that (i) this function is well
described by a power law over ranges in $\dot{M}_{acc}$ of up to at
least an order of magnitude and (ii) the corresponding power law
indices are well within the limits in Relation~(\ref{continuum})
(e.g. the index is $\simeq 0.7$ for both types of models for accretion
rates appropriate to CVs). 

Relation~(\ref{continuum}) establishes a connection between the expected
optical continuum flux and the accretion rate. This is the first
ingredient needed in a scaling relation between jet line EWs and
$\dot{M}_{acc}$. The second and final ingredient is a relation between
the expected jet line luminosity and the accretion rate.

To derive such a relation without having to make reference to the
relevant emission mechanism, we proceed in a very formal way. Noting
that all observed jet lines are Hydrogen or Helium recombination
lines, we may write quite generally  
\begin{equation}
L(line) = 4 \pi j_{line} \times EM.
\label{linelum0}
\end{equation}
In this relation, $j_{line}$ is the line emissivity corresponding to 
the relevant emission mechanism which, according to our working  
hypothesis, is the same in SSSs, T~Pyx and other CVs. In principle,
account should be taken of the fact that $j_{line}$ will be a 
function of the temperature and density in the emitting regions,
although any such dependence is likely to be minor compared to the
effect of the emission measure, 
\begin{equation}
EM = \int_{V}{n_e n_p dV} \simeq n_e^{2} V.
\label{em}
\end{equation}
Here, the integral is over the volume, $V$, of the line-emitting
region, and $n_e$, $n_p$ are the electron and proton number densities, 
respectively, in this region.

We now note that the electron density in the line-emitting region
must scale with the density and hence with the mass-loss rate, 
$\dot{M}_{jet}$, in the jet. 
Moreover, for a jet driven from an accretion disk we expect
$\dot{M}_{jet} 
\propto \dot{M}_{acc}$ to a first approximation, essentially
regardless of the jet driving mechanism (see e.g. Livio 1997a for
a discussion). We thus have \begin{equation}
L(line) \propto EM \propto n_e^2 \propto \dot{M}_{jet}^2 \propto
\dot{M}_{acc}^2,
\label{linelum}
\end{equation}
which is the desired relation between jet line luminosity and 
accretion rate. Note that scaling relations between {\em mass-loss}
rates and line luminosities have been used in studies of T~Tauri stars
(e.g. Hartigan, Edwards \& Ghandour 1995). In particular, the study by
Hartigan et al. (1995) shows that a clear correlation between 
$\dot{M}_{jet}$ and $\dot{M}_{acc}$ does exist in T~Tauri stars, in 
line with our assumption that $\dot{M}_{jet} \propto \dot{M}_{acc}$. 
However, it is also acknowledged that the correlation found by
Hartigan et al. (1995) exhibits substantial scatter. At the moment, 
it is not clear to what extent this corresponds to differences between
the parameters of individual systems (e.g. disk sizes), that could in
principle be corrected for (see below). 

Let us examine the fundamental step in the derivation of
Relation~(\ref{linelum}) -- the assumption of a direct proportionality
between mass loss and accretion rates -- in slightly more depth. Most
importantly, it should be kept in mind that this proportionality is
indeed an assumption. For example, if jets are driven by some form of
radiation pressure the luminosity of the central object will almost
certainly play a role in regulating the mass loss rate
\cite{proga2}.\footnote{This would probably remain true even if most
of the mass loss occurred relatively far out in the disk, since a 
very luminous central object will raise the disk's local surface 
temperature by irradiation. Thus from the point of view of a flow
driven by radiation pressure, a strongly irradiated disk is
equivalent to a disk with a higher mass transfer rate (see e.g. 
King 1997).} However, our fundamental working
hypothesis is that energy sources unrelated to the accretion process
are {\em not} needed for the formation of powerful
jets. In that context, the assumption that $\dot{M}_{jet} \propto
\dot{M}_{acc}$ is appropriate, as it corresponds to the
simplest, most literal interpretation of that hypothesis. (Turning the
argument around, we find, conversely, that jet driving by radiation
pressure is probably not consistent with our working hypothesis.) 
Some deviations from a precise proportionality between mass loss and
accretion rates are, of course, nevertheless possible. However, we
shall assume that they are smaller than the very conservative
uncertainties that we have allowed for in the dependence of optical
flux on accretion rate (Relation~\ref{continuum}).

Now, the equivalent width of an emission line is defined as 
\begin{equation}
EW(line) = \int_{line} \frac{(F_\lambda-F_c)}{F_c} d\lambda,
\label{ew}
\end{equation}
where the integral is over the line profile and $F_{\lambda}$, $F_{c}$
are, respectively, the total (line+continuum) and continuum fluxes at
a given wavelength. Similarly, the observed line luminosity can be
written as 
\begin{equation}
L(line) = 4 \pi d^2 \times \int_{line} (F_\lambda-F_c) d\lambda
\label{linelum2}
\end{equation}
where $d$ is the distance between the (isotropically) emitting source
and the observer. We can combine these two equations to give
\begin{equation}
L(line) = 4 \pi d^2 EW(line) F_c \propto EW(line) F_c,
\label{linelum3}
\end{equation}
where the continuum has been taken to be roughly constant over the
line profile. This is an excellent approximation for narrow spectral
features such as jet lines. 

The observed line luminosities in
Equations~(\ref{linelum2})~and~(\ref{linelum3}) can now be identified with
the intrinsic ones in Relations~(\ref{linelum0})~and~(\ref{linelum}), 
provided that (self-)absorption effects are unimportant. On
combining~(\ref{linelum3}) with~(\ref{continuum})~and~(\ref{linelum}),
we thus arrive at the simple scaling relation between jet line EW and
accretion rate that was our goal:
\begin{equation}
EW(line) \propto \frac{L(line)}{F_c} \propto
\frac{\dot{M}_{acc}^2}{\dot{M}_{acc}^{{\left({\frac{2}{3}}\right)}^{+1/3}_{-5/12}}}
\propto \dot{M}_{acc}^{{\left({\frac{4}{3}}\right)}^{+5/12}_{-1/3}}.
\label{final}
\end{equation}

We can improve on this relation in terms of its application
to different systems by considering the 
effects of other system parameters. One obvious factor that should be
taken into account is the binary inclination. If the optical continuum
is indeed dominated by radiation emitted by an optically thick disk,
the observed continuum flux will be affected by both foreshortening and 
limb-darkening. It will therefore scale as $\cos i ~ \eta (i)$, where 
$\eta(i)$ is the appropriate limb-darkening law. In the Eddington
approximation, which may be appropriate for CVs \cite{warner1},
$\eta(i)=\frac{1}{2}(1+\frac{3}{2}\cos i)$. Consequently, the RHS of 
Relation~(\ref{final}) should be multiplied by a factor of $\left[\cos i~
\eta (i)\right]^{-1}$ to account for inclination effects. 

In principle, the binary inclination will not affect the jet line
luminosity, so long as the line emitting volume is optically thin to 
recombination line photons. However, in practice 
the back side of the jet may be hidden from view by the optically
thick disk in very low inclination systems. In this case only a
blue-shifted satellite feature will appear in the spectrum, as seen 
in some T~Tauri stars (e.g. Edwards et al. 1987). Also, in very high
inclination systems, the line of sight velocity component of the
outflowing material may be too small for the jet line to appear well
separated from the line core.

A second parameter that should be taken into account is the disk size,
since this may set the scale of the jet cross-section, $A_{jet}$. 
As it stands, Relation~(\ref{final}) should only be applied 
when comparing objects for which these cross-sections are
approximately equal in the vicinity of the line-forming regions. This
is because, in a smooth flow, the emission measure will scale as
$EM \propto A_{jet}^{-1}$ ($n_e$ scales as $A_{jet}^{-1}$ and $V$ as
$A_{jet}$). In general, we might expect the size of the jet
cross-section to be set by some characteristic jet-formation radius,
$R_{jet}$, in the accretion disk, which must lie somewhere between the
radius of the object at the center of the accretion disk, $R_{WD}$,
and the full disk radius, $R_{disk}$. So unless $R_{jet} \simeq
R_{WD}$, Relation~(\ref{linelum}) should be modified to contain
an additional factor that accounts for possible differences in disk
radii between the objects being compared. 

To find a reasonable compromise between the two extreme possible
scalings of $R_{jet}$, with $R_{WD}$ on the one hand, and $R_{disk}$
on the other, we fleetingly appeal to a
specific physical picture for the origin of jets. We first note that 
the jet cross-section at some large distance, $l$, from the disk plane
will be proportional to $(\theta l)^2$ for any highly collimated jet
with approximately constant opening angle $\theta$. Now, if jets 
are collimated by a mechanism similar to poloidal collimation (e.g. 
Blandford 1993; Ostriker 1997), then we might expect that the 
distance down the jet to the line forming region scales as $l \propto 
R_{Alfven} \propto R_{disk}$ (where $R_{Alfven}$ is the Alfv\'{e}n
radius) and also that $\theta \propto
(R_{WD}/R_{disk})^{1/2}$ (e.g. Spruit 1996; Livio 1997a). 
We then have $A_{jet} \propto (\theta l)^2 \propto (R_{WD} R_{disk})$,
which implies that the relevant characteristic radius is $(R_{WD}
R_{disk})^{1/2}$, i.e. the geometric mean of the two limiting
radii. Guided by this, we adopt $R_{jet} \propto R_{disk}^{1/2 \pm
1/2}$, where the size of the ``error'' has been set so as to include
the most extreme plausible scalings.

Disk radii are relatively hard to measure observationally. In our
present application, it is therefore preferable to use Kepler's law to
recast any dependence 
on $R_{disk}$ into one on the orbital period, $P_{orb}$. This can be
done by noting that the binary separation, $a_{bin}$, is related to
$P_{orb}$ by $a_{bin} \propto P_{orb}^{2/3}$. Provided that the
mass ratios of the systems being compared are not too dissimilar,
their disk radii will be roughly equal fractions of their respective
binary separations and will therefore also scale as $P_{orb}^{2/3}$.
Thus the right hand side of Relation~(\ref{final}) should by
multiplied by a factor $A_{jet}^{-1} \propto R_{disk}^{-1 \pm 1}
\propto P_{orb}^{-2/3 \pm 2/3}$.

With the additional dependences on inclination and disk radius
included, Relation~(\ref{final}) now reads 
\begin{equation}
EW(line) \propto H~j_{line}~f_{fill}~
\frac{\dot{M}_{acc}^{{\left({\frac{4}{3}}\right)}^{+5/12}_{-1/3}}}{P_{orb}^{\frac{2}{3}\pm
\frac{2}{3}} \cos i ~\eta (i)}.
\label{final2}
\end{equation}
In this new relation, we have also made explicit the linear
scalings of the jet line EW with (i) $H$, the ``vertical'' scale
height of the line 
emitting region; (ii) $j_{line}$, the line emissivity (which will 
depend weakly on the density and somewhat more strongly on
the temperature in this region); (iii) $f_{fill}$, the filling factor, 
which accounts for the possibility that the flow is not smooth and 
the line forming region is not uniformly filled with emitting
gas. In principle, these could themselves depend indirectly on the 
accretion rate, and in ways that might be different for different 
dynamical models. Here, we take the plausible, but unproven view 
that any such dependencies are likely to be small compared to the
direct ones that we have accounted for. Based on our hypothesis that
SSSs, T~Pyx and other CVs all drive jets and share the same jet
driving and line emission mechanisms, we thus take $H$, $j_{line}$,
$f_{fill}$ to have similar values in all of these systems and ignore
these additional parameters in inter-comparisons. 

\section{Discussion}
\label{BLA}

\subsection{A prediction for the strength of jet lines in CVs}
\label{BLA2}

We are now in a position to use Relation~(\ref{final2}) to predict the
jet line EWs we would expect to see in ``ordinary'' CVs, according to
our working hypothesis. Since
T~Pyx {\em is} in fact a CV, its system parameters are more typical
of ``normal'' CVs than are those of the SSSs exhibiting jet
lines. It is therefore preferable to use T~Pyx as a
reference point in making predictions for other CVs, because the
ratios of the relevant factors in Relation~(\ref{final2}) will be closer
to unity and the associated uncertainties will be smaller. However, in
Section~\ref{BLA3} below we will check whether Relation~(\ref{final2})
is at least consistent with the observed accretion rate and EW ratios
of T~Pyx and the SSS RX~J0513-069. (See also the note at the end of
the manuscript, in which we show that the prediction derived in this
section by using T~Pyx as a reference datum is consistent with what 
is be obtained if RX~J0513-069 is used instead.)

Concerning T~Pyx, Webbink et al. (1987) give $\dot{M}_{acc}(T~Pyx)
\gtappeq 10^{-8}$~M$_{\sun}$~yr$^{-1}$ based on the short recurrence
time scale of its eruptions and $\dot{M}_{acc}(T~Pyx) \sim 5 \times
10^{-8}$~M$_{\sun}$~yr$^{-1}$ based on its optical colors. Here, we
adopt the latter estimate, which assumes that the optical light is 
due to the accretion disk, rather than to direct and/or 
reprocessed light from a hot WD. This is in line 
with our working hypothesis that a wind/energy source at disk center 
is not required to drive jets from accretion disks.
T~Pyx's orbital period is thought to be $P_{orb}(T~Pyx) \simeq
1.8$~hrs \cite{schaefer1}, placing the system below the 
period gap. As noted in Section~\ref{introduction}, the inclination
angle of T~Pyx is not well constrained observationally, although the
appearance of the satellite recombination lines themselves suggests a
high value $i(T~Pyx) \simeq 70^o$ if these features are formed in a well 
collimated jet. Finally, the equivalent widths of the H$\alpha$~jet
lines in T~Pyx can be measured from the data of Shahbaz et al. (1997)
and turn out to be about EW(T~Pyx) $\simeq 1$~\AA~on average, with the
strongest feature in any one of the observing epochs reaching about
twice this value (Shahbaz, private communication).

We now need to make some assumptions about the typical properties 
of ``ordinary'' CVs. The form of Relation~(\ref{final2}) shows that if
jets are present in these objects, the associated satellite
recombination lines are likely to be strongest in systems with 
high mass accretion rates, short orbital periods and high
inclinations (though not so high as to shift the jet lines into the
line core). Since it would be sufficient to falsify our working 
hypothesis for CVs with these properties, we adopt $\dot{M}_{acc}(CV)
\simeq 1 \times 10^{-8}$~M$_{\sun}$~yr$^{-1}$ (appropriate to NLs and
DNe in outburst), $P_{orb}(CV) \simeq P_{orb}(T~Pyx)$,
and $i(CV) \simeq i(T~Pyx)$. (We note in passing that there are actually
no well-established non-magnetic NL variables with periods shorter than
3.2~hrs, although there is a fair number of DNe with $P_{orb} \leq
2$~hrs.) 

We can now use Relation~(\ref{final2}) to predict the jet line EWs 
we expect to see in this most favorable sub-group of ``ordinary'' CVs,
according to our working hypothesis. To this end, we take the ratio of the
two separate relations (one for the normal CVs, one for T~Pyx), solve
for $EW(CV)$ and substitute our adopted parameters. This yields
$EW(CV) \simeq 0.1^{+0.1}_{-0.04}$~\AA. While it may be possible to
increase the upper limit implied by this result somewhat -- by taking
$P_{orb}(CV)$ to be shorter or $i(CV)$ to be higher, for example -- it
is clear that CV jet lines, if they exist, would be at best marginally
detectable in typical optical spectra. As a result, we are forced to
conclude that our working hypothesis and its corollary -- that a hot 
central object is inessential to the formation of jets and that CVs do
in fact drive jets -- {\em cannot} yet be ruled out. 

Let us take a step back at this point to make it clear what we are --
and are not -- claiming. We started by adopting the working hypothesis 
that the formation of powerful jets does {\em not} require the
presence of an additional energy source at disk center. As a
corollary, we assumed that ``ordinary'' CVs harbor the same kind of
jets that may be present in T~Pyx (as indicated by the satellite
recombination lines that are observed in that object). We then showed
that based on these assumptions one can derive a simple scaling law 
which can be used to predict the expected strength 
of these jet lines in the optical spectra of ordinary
CVs. The predicted jet line EWs for these systems turned out be very 
small, even for objects with nearly optimal system parameters. We
therefore concluded that the lack of jet line detections 
in the optical spectra of ordinary CVs is not yet in conflict with our
working hypothesis, i.e. that jets {\em may} be present in ordinary
CVs. Note that we do {\em not} claim to have shown that ordinary 
CVs actually {\em do} contain jets. After all, an inability to falsify a
hypothesis does not prove it. Summarized succinctly, our conclusion is
that {\em the non-detection of jet lines in existing optical spectra
of ``ordinary'' CVs should not yet be taken to imply that these
systems cannot harbor collimated jets}.

Two further points need to be made regarding this statement. First,
even though we have been unable to rule out the presence of jets in
``ordinary'' CVs on the basis of existing data, the predicted EWs of a
few hundredths up to a few tenths of Angstroms may not be beyond the
reach of high resolution, high signal-to-noise optical spectra. Thus we
strongly encourage observers to search for the signatures of jets in
the spectra of appropriately selected CVs.

Second, it was assumed above that T~Pyx's optical continuum 
is dominated by the radiation field emitted by a standard accretion
disk. However, if (intermittent) nuclear burning really does take place 
in T~Pyx, the surface of the WD at the center of the disk will be
extremely hot. It is therefore worth asking whether (some of) the
optical continuum could actually be direct or reprocessed radiation
emitted by the WD, and what effect this may have on our conclusions. 

A numerical calculation similar to that described following
Relation~(\ref{continuum1}) in Section~\ref{scale} shows that direct
light from the WD is unlikely to be of any importance, even if the 
temperature of the WD is as high as a few times $10^5$~K, and the
accretion rates as low as $10^{-8}$~M$_{\sun}$~yr$^{-1}$. 
To judge the potential significance of reprocessed WD radiation, we
rely on the recent work of King (1997), who derived a 
simple condition that can be used to estimate the relative importance
of dissipation and reprocessing in a CV accretion disk. More
specifically, King (1997) showed that reprocessing of WD radiation 
will begin to have a dominant effect
on the local disk temperature if $L_{WD} \gtappeq 2.5 L_{acc}
(1-\beta)^{-1}$, where $L_{WD}=4\pi R_{WD}^2 \sigma T_{WD}^4$ and
$L_{acc} = GM_{WD} \dot{M}_{acc}/R_{WD}$ are the WD and total
accretion luminosities, 
respectively, and $\beta$ is the albedo of the disk surface. To give a
numerical example, we note that if reprocessing is assumed to be
efficient ($\beta \simeq 0$), the temperature distribution in a disk
around a $1M_{\sun}$~WD accreting at a rate of $\dot{M}_{acc} =
10^{-8} M_{\sun}$~yr$^{-1}$ will be dominated by reprocessing
if $T_{WD} \gtappeq 2 \times 10^5$~K.

If reprocessed WD radiation is in fact contributing significantly to 
T~Pyx's optical continuum, then our previous prediction for the
strength of jet lines in other CVs no longer applies, since our
continuum scaling law, Relation~(\ref{continuum}), ceases to be
valid. Qualitatively, the effect of this will be to increase the predicted 
EWs significantly, since (a) the adopted accretion rate
for T~Pyx is almost certainly an overestimate in this case, and (b)
the extra contribution to the continuum that is ultimately due to
nuclear burning on the WD (and not to accretion) is making the jet
lines appear weaker than if only the disk were producing
the continuum. Quantitatively, these effects can be corrected for 
by multiplying the predicted EWs by a factor of $f_{\dot{M}}^{2}$,
where $f_{\dot{M}}>1$ is the factor by which T~Pyx's accretion rate
has been overestimated. The dependence on $f_{\dot{M}}$ {\em squared} 
arises because the part of correction (a) that is related to the
scaling of the continuum flux with accretion rate exactly cancels
correction (b). This leaves the scaling of the line luminosity with 
accretion rate as the only relevant factor.

It is now easy to see that if irradiation is very important in T~Pyx and
has caused us to overestimate the accretion rate by a significant
amount, then the non-detection of jet lines in the spectra of other
CVs does become inconsistent with the presence of jets in these
systems. Indeed, if $f_{\dot{M}}\gtappeq 4$, then even the previously
derived lower limit of 0.06~\AA~on the jet line EWs in (suitably
selected) CVs becomes as large as 1~\AA~and thus comparable to the
strength of the same features in T~Pyx. In practical terms, this means
that studies of T~Pyx aimed at deriving $T_{WD}$ (or, more precisely, 
$L_{WD}/L_{acc}$) for this system may provide yet another way to
falsify our working hypothesis observationally in the future.

\subsection{The scaling relation applied to T~Pyx and the SSSs}
\label{BLA3}

Given that the jets in T~Pyx and the SSSs are presumably of the same
type, it is natural to try and use these systems to check our scaling
relation for the jet line EWs. Unfortunately, 
the accretion rates of the relevant SSSs are only poorly constrained
and, in addition, disk irradiation by the hot WD is likely
to be very strong in the SSSs. As a consequence, a rigorous test
of Relation~(\ref{final2}) via this route is not possible. However,
we will nevertheless proceed to apply our scaling relation to T~Pyx
and the SSS RX~J0513-069, partly to illustrate these problems, and
partly to perform at least a rough consistency check.

In their study of RX~J0513-069, Southwell et al. (1996) state that
$\dot{M}_{acc} \sim 10^{-5}$~M$_{\sun}$~yr$^{-1}$ is required if the
optical luminosity of this system is to be ascribed entirely to a
standard accretion disk. An accretion rate this high is of the order
of the Eddington value, and Southwell et al. (1996) therefore conclude
that it is almost certainly an overestimate. They argue that
irradiation of the disk and 
secondary star, as well as (perhaps) direct light from the hot WD are
likely to contribute significantly to the optical light. Consequently,
they prefer a lower value of about $10^{-6}$~M$_{\sun}$~yr$^{-1}$ for
the accretion rate. To make progress in the face of this uncertainty,
we will adopt the higher value to start with and then check {\em a
posteriori} what value this implies for the correction factor
$f_{\dot{M}}^2$. Regarding RX~J0513-069's other relevant parameters, 
Southwell et al. (1996) give values of $P_{orb} \simeq 18$~hrs for the
orbital period, and, based on the mass function of the system, $i
\simeq 10^o$ for the inclination.

Adopting these parameters for RX~J0513-069, and using the same
parameters as above for T~Pyx, we would predict a best-bet ratio for
the EWs of the jet lines in these two systems of about 50 (in favor
of the SSS). Now, Southwell et al. (1996) measure the equivalent
widths of the blue and red H$\alpha$~jet satellite lines in
RX~J0513-069 to be 
EW(SSS,blue) $\simeq 1.6$~\AA~and EW(SSS,red) $\simeq 2.6$~\AA,
respectively. Thus the actual ratio of the jet line EWs in
RX~J0513-069 and T~Pyx is only about 2. If we interpret this as a
result of disk irradiation in the SSS, then the 
correction factor $f_{\dot{M}}^2 \simeq 25$ and $f_{\dot{M}} \simeq
5$. Consequently, we would predict the true accretion rate in
RX~J0513-069 to be about $\dot{M}_{acc} \sim 2 \times
10^{-6}$~M$_{\sun}$~yr$^{-1}$, which is in line with the value of
$10^{-6}$~M$_{\sun}$~yr$^{-1}$ preferred by Southwell et al. (1996). 
We do not attach too much weight to this apparent consistency, because
there are large observational uncertainties associated with the
ratios constructed from two of the relevant parameters (accretion rate
and inclination). Moreover, the accretion rate and orbital period
ratios are so large for these systems that the theoretical
uncertainties expressed by the ``errors'' in Relation~(\ref{final2})
also become rather large.

It is finally interesting to consider briefly the implications of
adopting the complement of our working hypothesis. Specifically, 
we may ask 
whether a consistent physical picture capable of accounting for the
relative strengths of the jet lines in T~Pyx and RX~J0513-069 can
also be found if we assume that a hot, central object is in fact 
present in both systems and is crucial for driving the observed
jets. To answer this question, 
we take $\dot{M}_{jet} \propto L_{WD}$ and assume the extreme
case of $L_{WD} >> L_{acc}$. The disk is then still 
quite likely to dominate the optical flux (c.f. the numerical 
estimates for the direct WD contribution given previously), 
but its local temperature distribution will be dominated by
irradiation, not dissipation (see Section~\ref{BLA2}). 
Since the disk will be extremely hot
in this case, we may further assume that the optical
waveband lies on the Rayleigh-Jeans tail of the disk spectrum now,
i.e. 
$F_{opt} \propto L_{disk}^{1/4} \propto L_{WD}^{1/4}$ (the latter
holds since $L_{disk}$ is now dominated by $L_{WD}$). We can then
replace the dependence on $\dot{M}_{acc}$ in Relation~(\ref{final2})
with one on 
$L_{WD}$, giving $EW(line) \propto L_{WD}^{7/4} \propto 
T_{WD}^7$. Adopting again an EW ratio of 2 for RX~J0513-069 and 
T~Pyx, we find that $T_{WD}(SSS) \simeq 2~T_{WD}(T~Pyx)$ in this
simplistic picture, if the remaining parameter dependences in
Relation~(\ref{final2}) are assumed to stay unchanged. 

This reasonable looking result should of course not be taken too
seriously. However, the moral of this simple calculation is that it is
certainly possible to account for the jet line EW differences between
T~Pyx and RX~J0513-069 in the context of a model in which the presence
of an 
energy source at disk center {\em is} a crucial ingredient in
driving the observed jets. This prompts us to stress again
that our analysis in this paper has only shown that the presence of
jets in CVs should not be ruled out simply because no jet lines have
so far been detected in the optical spectra of these systems. We have
by no means demonstrated that jets are actually present, or are even
likely to be present, in ordinary CVs.

{\bf Note added:} After this paper was accepted for publication, we received
a draft of a work by Margon \& Deutsch, in which it is argued that the
satellite lines seen in T~Pyx are in fact due to [N~{\sc
ii}]~$\lambda\lambda$6548,6584 and are formed in the complex
velocity field of T~Pyx's nova shell(s). While the analytic scaling
relation we derived in Section~2 retains its validity (and, we
believe, usefulness) if this interpretation turns out to be correct,
the same is not true for the prediction we made for the jet line EWs
in ordinary CVs (since this is based on the assumption that
T~Pyx's satellite lines are jet features). The best we can do in this
case is to derive a new prediction by scaling down directly from
one of the SSSs to CVs. To do this, 
we use Southwell~et al.'s (1997) inclination, orbital period and
accretion rate estimates for RX~J0513-069 ($i\simeq 10^o$; $P_{orb}
\simeq 18$~hrs; $\dot{M}_{acc}(apparent)\sim
10^{-5}$~M$_{\sun}$~yr$^{-1}$ with $f_{\dot{M}}=10$) and, as before,
parameters appropriate to an optimally selected, ``ordinary'' CV
($i\simeq 70^o$; $P_{orb} \simeq 1.8$~hrs; $\dot{M}_{acc} \sim
10^{-8}$~M$_{\sun}$~yr$^{-1}$). Ignoring limb-darkening ($\eta(i)=1$),
we obtain a new prediction of $EW(CV) \sim 0.3$~\AA. Even though the 
uncertainties on this number are substantial and hard to quantify (see
Section~2.3), this estimate still suggests it would be premature to 
rule out the presence of jets in CVs completely at this stage. 
\footnote{Note that if jets are present in CVs but jet lines are not
seen in T~Pyx, the {\em absence} of the latter would have to be
attributed to one or both of the 
following: (i) T~Pyx's inclination is much lower than $70^o$; (ii)
irradiation is increasing the brightness of the accretion disk in
T~Pyx substantially.} We therefore suggest that an optical survey of
suitably selected CVs to search for jet lines is called for,
regardless of the nature of the satellite lines in T~Pyx.

\section*{Acknowledgments}
We are grateful to Tariq Shahbaz for providing us with measurements of
the satellite line EWs in T~Pyx and to the referee, Scott Kenyon, for
a concise and useful report. C.K. would also like to thank Mike
Goad for useful discussions regarding the formation of jet lines and
Knox Long for financial support through a postdoctoral fellowship at
STScI. M.L. was supported by NASA Grant NAGW-2678.

\label{lastpage}
 

\begin{thebibliography}{99}

\bibitem[\protect\citename{Baptista}{~1995}]
{bap}
Baptista, R., Horne, K., Hilditch, R. W., Mason, K. O. \& Drew,
J. E. 1995, ApJ, 448, 395

\bibitem[\protect\citename{Becker, Remillard, \& Rappaport}{~1997}]
{becker1}
Becker C.~M., Remillard R.~A.,  Rappaport S.~A., 1997, in preparation

\bibitem[\protect\citename{Blandford}{~1993}]
{bland3}
Blandford R.~D., 1993, in Burgarella D., Livio M.,  O'Dea C.~P., ed,
  Astrophysical Jets.
\newblock Cambridge University Press, Cambridge, p. 263

\bibitem[\protect\citename{Crampton et~al.}{~1987}]
{crampton1}
Crampton D., Cowley A.~P., Hutchings J.~B., Schmidtke P.~C.,  Thompson I.~B.,
  1987, ApJ, 321, 745

\bibitem[\protect\citename{Crampton et~al.}{~1996}]
{crampton2}
Crampton D., Hutchings J.~B., Cowley A.~P., Schmidtke P.~C., McGrath T.~K.,
  O'Dononghue D.,  Harrop-Allin M.~K., 1996, ApJ, 456, 320

\bibitem[\protect\citename{Drew}{~1991}]
{drew3}
Drew J.~E., 1991, in Bertout C., Collin-Souffrin S., Lasota J.~P.,  Thran
  Thran~Van J., ed, Structure and Emission Properties of Accretion Disks.
\newblock Editions Frontiers, Paris, p. 331

\bibitem[\protect\citename{Edwards et~al.}{~1987}]
{edwards1}
Edwards S., Cabrit S., Strom S.~E., Heger K.~M., I.and~Strom,  Anderson E.,
  1987, ApJ, 321, 473

\bibitem[\protect\citename{Hartigan}{1995}]
{hartigan1}
Hartigan, P., Edwards, S. \& Ghandour, L. 1995, ApJ, 452, 736

\bibitem[\protect\citename{Haswell et~al.}{~1994}]
{haswell1}
Haswell C.~A., Horne K., Thomas G., Patterson J.,  Thorstensen J.~R., 1994, in
  Shafter A.~W., ed, Interacting Binary Stars.
\newblock ASP, San Francisco, p. 268

\bibitem[\protect\citename{King}{~1997}]
{king4}
King A.~R., 1997, MNRAS, 288, L16

\bibitem[\protect\citename{Horne}{~1995}]
{horne1}
Horne, K. \& Stiening, R. F 1985, MNRAS, 216,933

\bibitem[\protect\citename{Knigge \& Drew}{~1997}]
{me6}
Knigge C.,  Drew J.~E., 1997, ApJ, in press

\bibitem[\protect\citename{Knigge et~al.}{~1997}]
{me5}
Knigge C., Long K.~S., Blair W.~P.,  Wade R.~A., 1997, ApJ, 476, 291

\bibitem[\protect\citename{Knigge et~al.}{~1998}]
{me7}
Knigge C., Long K.~S., Wade, R. A., Baptista, R., Horne, K., Hubeny,
I., Rutten, R. G. M. 1998, ApJ, 499, in press

\bibitem[\protect\citename{Livio}{~1997a}]
{livio4}
Livio M., 1997a, in Wickramasinghe D.~T., Bicknell G.~V.,  Ferrario L., ed,
  {Accretion Phenomena and Related Outflows}.
\newblock ASP Conference Series, p.~8

\bibitem[\protect\citename{Livio}{~1997b}]
{livio5}
Livio M., 1997b, in 13th North American Workshop on Catalysmic Variables, p. in
  press

\bibitem[\protect\citename{Ostriker}{~1997}]
{ostriker1}
Ostriker E.~C., 1997, in Wickramasinghe D.~T., Ferrario L.,  Bicknell G.~V.,
  ed, {Accretion Phenomena and Related Outflows}.
\newblock ASP Conference Series, p. in press

\bibitem[\protect\citename{Proga, Stone, \& Drew}{~1997}]
{proga2}
Proga D., Stone J.~M.,  Drew J.~E., 1997, MNRAS, in press

\bibitem[\protect\citename{rutten}{~1992}]
{rutten}
Rutten, R. G. M., van Paradijs, J. \&  Tinbergen, J. 1992, A\&A, 260, 213

\bibitem[\protect\citename{Schaefer et~al.}{~1992}]
{schaefer1}
Schaefer B.~E., Landolt A.~U., Vogt, N., Buckley D., Warner B., Walker
A.~R., Bond H.~E., 1992, ApJS, 81, 321

\bibitem[\protect\citename{Seitter}{~1984}]
{seitter1}
Seitter W.~C., 1984, Ap\&SS, 99, 95

\bibitem[\protect\citename{Shahbaz et~al.}{~1997}]
{shahbaz1}
Shahbaz T., Livio M., Southwell K.~A.,  Charles P.~A., 1997, ApJ, in press

\bibitem[\protect\citename{Shlosman, Vitello, \& Mauche}{~1996}]
{shlos2}
Shlosman I., Vitello P.~A.~J.,  Mauche C.~W., 1996, ApJ, 461, 377

\bibitem[\protect\citename{Southwell, Livio, \& Charles}{~1997}]
{southwell2}
Southwell K.~A., Livio M.,  Charles P.~A., 1997, MNRAS, submitted

\bibitem[\protect\citename{Southwell et~al.}{~1996}]
{southwell1}
Southwell K.~A., Livio M., Charles P.~A., O'Donoghue D.,  Sutherland W.~J.,
  1996, ApJ, 470, 1065

\bibitem[\protect\citename{Spruit}{~1996}]
{spruit2}
Spruit H.~C., 1996, in Wijers R.~A. M.~J., Davies M.~B.,  Tout C.~A., ed,
  {Evolutionary Processes in Binary Stars}.
\newblock Cambridge University Press, Cambbridge, p. 249

\bibitem[\protect\citename{Szkody}{~1995}]
{szko4}
Szkody P., 1995, in Buckley D.~A.~H.,  Warner B., ed, Cape Workshop on Magnetic
  Cataclysmic Variables.
\newblock ASP Conference Series Vol. 85, San Francisco, p.~54

\bibitem[\protect\citename{van~den Heuvel et~al.}{~1992}]
{heuvel1}
van~den Heuvel E.~P.~J., Bhattacharya D., Nomoto K.,  Rappaport S.~A., 1992,
  A\&A, 262, 97

\bibitem[\protect\citename{Vermeulen}{~1993}]
{vermeulen1}
Vermeulen R., 1993, in Burgarella D., Livio M.,  O'Dea C.~P., ed, Astrophysical
  Jets.
\newblock Cambridge University Press, Cambridge, p. 241

\bibitem[\protect\citename{Warner}{~1986}]
{warner1}
Warner B., 1986, MNRAS, 222, 11

\bibitem[\protect\citename{Webbink et~al.}{~1987}]
{webbink1}
Webbink R.~F., Livio M., Truran J.~W.,  Orio M., 1987, ApJ, 314, 653

\end{thebibliography}
\end{document}